\documentclass[twocolumn,prl,aps,superscriptaddress,reprint]{revtex4-1}

\usepackage{mathptmx}
\usepackage{subfigure}
\usepackage{psfrag,graphicx}
\usepackage{pict2e}
\usepackage{dcolumn}
\usepackage{bm}
\usepackage{color}
\usepackage{latexsym}
\usepackage{epstopdf}
\usepackage{color}
\usepackage[english]{babel}
\usepackage{latexsym}
\usepackage{amsmath}
\usepackage{amssymb}
\usepackage{amsfonts}
\usepackage{bm}
\usepackage{natbib}
\usepackage{appendix}
\usepackage{simplewick}
\definecolor{myblue}{rgb}{0.2,0.2,0.8}
\definecolor{myred}{rgb}{1,0.,0.3}
\usepackage{xr}
\usepackage[colorlinks=true,citecolor=myblue,linkcolor=myred]{hyperref}

\begin{document}
\title{Topological edge states in periodically-driven trapped-ion chains}

\author{Pedro Nevado}
\email{pedro.nevado.serrano@gmail.com}
\affiliation{Department of Physics and Astronomy, University of Sussex, Brighton BN1 9QH, United Kingdom}

\author{Samuel Fern\'andez-Lorenzo}
\affiliation{Department of Physics and Astronomy, University of Sussex, Brighton BN1 9QH, United Kingdom}

\author{Diego Porras}
\email{D.Porras@Sussex.ac.uk}
\affiliation{Department of Physics and Astronomy, University of Sussex, Brighton BN1 9QH, United Kingdom}

\date{\today}

\begin{abstract}
Topological insulating phases are primarily associated with condensed-matter systems, which typically feature short-range interactions. Nevertheless, many realizations of quantum matter can exhibit long-range interactions, and it is still largely unknown the effect that these latter may exert upon the topological phases. In this Letter, we investigate the Su-Schrieffer-Heeger topological insulator in the presence of long-range interactions. We show that this model can be readily realized in quantum simulators with trapped ions by means of a periodic driving. Our results indicate that the localization of the associated edge states is enhanced by the long-range interactions, and that the localized components survive within the ground state of the model. These effects could be easily confirmed in current state-of-the-art experimental implementations.
\end{abstract}

\maketitle

\emph{Introduction.-} Topological phases are one of the most exotic forms of quantum matter. Among their many intriguing traits, we find that they are robust against local decoherence processes, or feature fractional particle excitations with prospective applications in quantum information processing \cite{Kitaev2003AoP,Nayak2008RMP}. Some of the simplest systems showcasing non-trivial topological order are the topological insulators \cite{Hasan2010RMP,Moore2010N,Qi2011RMP,Bernevig2013}, gapped phases of non-interacting fermions which present gapless edge states. Despite of several experimental realizations \cite{Koenig2007S,Chen2009S}, the preparation and measurement of topological insulators is typically difficult in the solid state. Analog quantum simulators \cite{Bloch2008RMP,Lewenstein2007AiP,Porras2004PRLa,Friedenauer2008NP,Schneider2012RoPiP,Jurcevic2014N,Richerme2014N}, on the other hand, offer the possibility of exploring and exploiting the topological insulating phases, because of their inherent high degree of controllability. Furthermore, interactions in a quantum simulator can be tuned at will, opening up the possibility of investigating new regimes of the underlying models. 

Topological edge states usually occur in the insulating phase as long as an associated bulk invariant attains a non-trivial value, and the generic symmetries of the underlying Hamiltonian are preserved \cite{Ryu2002PRL}. This property --known as the bulk-edge correspondence-- is a generic feature of topological insulators. However, if interactions are taken into account, the presence of edge states is no longer guaranteed. For instance, it has been shown that one of the edge states present in the Mott insulating phase of the Bose-Hubbard model on a 1D superlattice is not stable against tunneling \cite{Grusdt2013PRL}. In this work, we extend these considerations to the case of interactions which are explicitly long ranged. Since topological phases are characteristically robust against \emph{local} perturbations, but long-range interactions may not qualify as such, there is an ongoing effort to elucidate their effect upon the topological states \cite{Gong2016PRB,Patrick2016Ae,Bermudez2017PRB}. This question is not of exclusive theoretical interest, since many experimental systems implementing topological phases of matter feature long-range interactions. In particular, we will show that trapped-ion quantum simulators can realize a long-range interacting version of one of the simplest instances of a topological insulator, the Su-Schrieffer-Heeger (SSH) model \cite{Su1979PRL,Su1980PRB,Heeger1988RMP}
\begin{equation}
H_{\rm SSH} = J\sum_{j=1}^{N-1} \left(1+\delta(-1)^j\right)\left(\sigma^+_j\sigma^-_{j+1} +{\rm H.c.}\right).
\label{h.ssh.original}
\end{equation}
The SSH model presents topological edges states for $\delta>0$, which, e.g., near the left end of the chain are of the form $|{\rm E.S.}\rangle \sim \sum_{j=1}^N e^{(N-j+1)/\xi_{\rm loc}}\sigma^+_j\left|\downarrow \downarrow \downarrow \ldots \right\rangle$, where the localization length can be related to the dimerization $\delta$ through \cite{CamposVenuti2007PRA}
\begin{equation}
\xi_{\rm loc} = -2/\ln \frac{1-\delta}{1+\delta},\quad0<\delta<1.
\label{loc.length.discrete}
\end{equation}
The addition of long-range inter-ion couplings on (\ref{h.ssh.original}) turns this model into a highly non-trivial interacting problem. However, we will show that, owing to the single-particle addressability available in trapped-ion setups, the edge states can be studied as one-body solutions, and that their properties survive when interactions are taken into account. 

This Letter is structured as follows. (i) We begin showing how to implement the interacting SSH model with trapped-ion quantum matter. (ii) We then study its one-excitation subspace, and locate the topological phase. (iii) We perform an effective description of the low-energy sector, and establish the dependence of the localization length with the range of the interactions. Also, we provide a protocol for the detection of the edge states. (iv) Finally, we study the correlations in the ground state, and establish the survival of the boundary modes against interactions.

\emph{Realization of the spin SSH Hamiltonian.-}  
We consider a set of $N$ trapped ions arranged along a 1D chain. Two optical or hyperfine 
levels $\left|\uparrow\right\rangle,\left|\downarrow\right\rangle$ encode an effective spin, 
such that
$\left|\uparrow\right\rangle 
\left\langle\uparrow\right|-\left|\downarrow\right\rangle 
\left\langle\downarrow\right|\equiv\sigma^z$ \cite{Leibfried2003RMP}.
The vibrations of the chain can be approximated by a set of harmonic modes, 
$H_{\rm ph} = \sum_{n = 1}^N \omega_n a^\dagger_n a_n$. We add a state-dependent force conditional on the internal states of the ions \cite{Sorensen1999PRL,Solano1999PRA,Sackett2000N,Lee2005JOB}, whose frequency is fairly off-resonant with any motional excitation,
\begin{equation}
H_{\rm f}(t) = g \sum_{j,n  = 1}^N \sigma^x_j \left( M_{j,n} e^{i \delta_n t} a^\dagger_n + {\rm H.c.} \right).
\label{H.force}
\end{equation}
$\delta_n = \omega_n - \Delta \omega$, where $\Delta \omega$ is the laser detuning  
with respect to the internal level and $M_{j,n}$ is the phonon wave-function. We 
assume that the force acts in the direction transverse to the ion chain. In this case the 
mode $n = N/2$ has the minimum energy. After tracing out the vibrational bath, the dynamics 
of the spins can be described by an Ising Hamiltonian \cite{Porras2004PRLa}
\begin{equation}
H_{\rm Ising} = \sum_{j,l=1}^N J_{j,l}^{(\rm ions)} \sigma^x_j \sigma^x_l + \frac{\Omega}{2}\sum_{j=1}^N \sigma^z_j,
\label{hamiltonian.ising}
\end{equation}
where the extra transversal field, $\Omega$ can be realized with a microwave or a Raman transition. 
The nature of the couplings $J_{j,l}^{(\rm ions)}$ depends on the width of dispersion relation of the motional modes, $t_{\rm C}$, and the detuning of the laser from the bottom of the band, $\delta_{N/2}$ \cite{Deng2005PRA}. 
All through this work we assume $\delta_{N/2} > 0$. In \cite{Nevado2016PRA} we showed that depending on the relative values of $\delta_{N/2}$ and $t_{\rm C}$, we can distinguish two regimes:
(i) long-range limit $(\delta_{N/2}\ll t_{\rm C})$, in which $J_{j,l}^{(\rm ions)}\sim e^{-|j-l|/\xi_{\rm int}}$. The spin coupling takes a Yukawa-like form, with 
$\xi_{\rm int} = \sqrt{\log(2)/2} \sqrt{t_{\rm C}/\delta_{N/2}}$, 
 and
(ii) short-range limit $(\delta_{N/2}\gg t_{\rm C})$, in which the couplings decay as $\sim |j-l|^{-3}$, and the interactions are effectively among nearest neighbors only.

Since $\sigma^x_j = \sigma^+_j + \sigma^-_j$, Hamiltonian (\ref{hamiltonian.ising}) contains terms of the form $\sigma^+_j \sigma^+_l, \sigma^-_j\sigma^-_l$, which do not occur in (\ref{h.ssh.original}). To eliminate these we assume a rotating wave approximation in the limit $\Omega \gg J_{j,l}$. To obtain the SSH model we consider driving the chain with a time-dependent field. 
Periodic drivings are known to render effective Hamiltonians in which specific terms can be adiabatically eliminated, and the interactions are non-trivially dressed \cite{Fernandez-Lorenzo2016NJoP}. 
In our case this dressing must also contain some spatial structure to give rise to the periodicity of the couplings in the SSH model. 
We exploit the possibility of globally imprinting inhomogeneous couplings upon the chain, by 
taking advantage of the optical phase of the laser fields \cite{Nevado2016PRA},
\begin{equation}
H_{\rm driving}=\frac{\eta \omega_{\rm d}}{2} \cos(\omega_{\rm d}t) \sum_{j=1}^N\cos(\Delta k d_0 j +\phi)\sigma^z_j.
\label{H.driving}
\end{equation}
This driving relies on a standing wave modulated in time with frequency $\omega_{\rm d}\ll\Omega$, which should be implemented by a different set of lasers than the state-dependent force in Eq. (\ref{H.force}). $\eta$ is the (dimensionless) coupling strength, $\Delta k$ is the wave vector along the chain axis, and $\phi$ is a global optical shift. 
We assume that the ions are equally spaced by $d_0$, so their equilibrium positions are $r_j^{(0)}=d_0 j$. This is a good approximation in the center of a Coulomb crystal in a RF trap \cite{Paul1990RMP}, or describes a linear array of microtraps \cite{Chiaverini2005QIC,Seidelin2006PRL,Home2009S}. 
Now we move into a rotating frame such that $H_{\rm Ising} + H_{\rm driving}\equiv H_{\rm total} \to {H}_{\rm total}'$, with ${H}_{\rm total}' = U(t)H_{\rm total} U^{\dagger}(t) - i U(t) \frac{d}{dt} U^{\dagger}(t)$, $U(t) = \exp{[i \sum_{j=1}^N \Delta_j(t) \sigma^z_j]}$, and
\begin{equation}
\Delta_j(t) = \frac{\Omega}{2} t + \frac{\eta \omega_{\rm d}}{2}  \cos(\Delta k d_0 j +\phi)\int_0^t \cos(\omega_{\rm d} t') dt'.
\end{equation}
The condition $\max_{j,l}|J_{j,l}^{(\rm ions)}|\ll \Omega$ ensures that the anomalous terms are fast rotating, whereas those that preserve the $z-$component of the spin are renormalized by the phases $e^{\pm i(\Delta_j(t)-\Delta_l(t))}$. These quantities can be simplified by using suitable trigonometric identities along with the Jacobi-Anger expansion $e^{iz\sin\theta}=\sum_{n=-\infty}^{\infty} B_n(z)e^{in\theta}$, where $B_n(z)$ are the Bessel functions of the first kind \cite{Weber2004}. Assuming that $\omega_{\rm d}\gg \max_{j,l}|J_{j,l}^{(\rm ions)}|$, the only non-fast-rotating contribution comes from $n=0$, and ${H}_{\rm total}'\simeq H_{\rm SSH}^{(\rm ions)}$, with
\begin{equation}
H_{\rm SSH}^{(\rm ions)} =  \sum_{j,l=1}^N J_{j,l}^{(\rm ions)} \mathcal{J}_{j,l}^{\pi/2} \left(\sigma^+_j \sigma^-_l+\sigma^-_j \sigma^+_l\right),
\label{hamiltonian.ssh.spins.long.range}
\end{equation}
where we fix $\Delta kd_0 =\pi/2$ to achieve the periodic couplings
\begin{equation}
\mathcal{J}_{j,l}^{\pi/2} = B_{0} \left(2\eta\sin\left(\frac{\pi }{4}(j+l)+\phi\right)\sin\frac{\pi}{4}(j-l)\right).
\label{bessel.couplings}
\end{equation}
Since $\mathcal{J}_{j,j+1}^{\pi/2} = \mathcal{J}_{j+T,j+T+1}^{\pi/2}$ with $T=2$, these couplings reproduce the dimerization of the original SSH model in the limit of nearest-neighbor interactions. We will refer to the spin implementation (\ref{hamiltonian.ssh.spins.long.range}) as the generalized SSH model. In analogy with (\ref{h.ssh.original}), the dimerization is given by the differential ratio of the couplings between sites with $j$ even and odd, i.e.,
\begin{equation}
\delta = \frac{\mathcal{J}_{2,3}^{\pi/2}-\mathcal{J}_{1,2}^{\pi/2}}{\mathcal{J}_{2,3}^{\pi/2}+\mathcal{J}_{1,2}^{\pi/2}}.
\label{generalized.dimerization}
\end{equation}
In Eq. (\ref{generalized.dimerization}), $J^{\rm (ions)}_{j,l}$ factors out, since $J^{\rm (ions)}_{j,l}=J^{\rm (ions)}_{j-l}$.

Finally, we remark that we can easily extend our derivation to account for the effect of an inhomogeneous ion-ion spacing, whose main effect would be to induce an extra site-dependence in the couplings to the standing-wave. Since the topological properties investigated below are robust against perturbations, we can expect our results to be valid even when small inhomogeneities are considered.

\begin{figure}[h!]{
		\centering
		\includegraphics[width=1\linewidth]{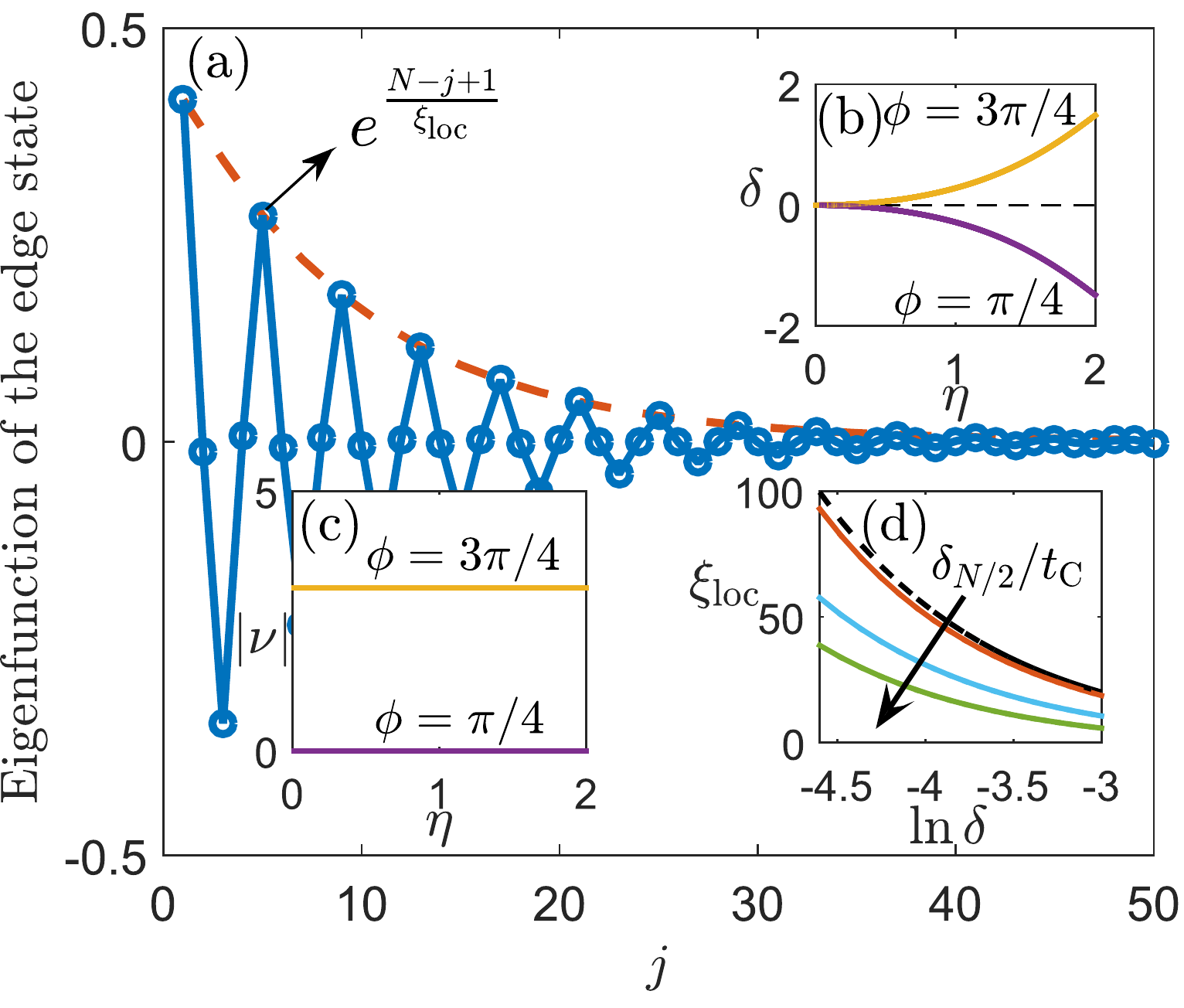}
		\caption{(color online) (a) Plot of the mid-gap state (circles) near the left end, with $\delta_{N/2}/t_{\rm C} = 4$, for $N=100$, $\delta\simeq0.1$ ($\phi = 3\pi/4$, $\eta\simeq0.62$). The solid line is a guide-for-the-eye, and the dashed is the envelope of the edge state. (b) Dimerization as a function of $\eta$ in the chiral limits. (c) Zak phase for different values of $\eta$, signaling the topologically trivial ($|\nu|=0$) and non-trivial ($|\nu|=\pi$) phases. (d) Dependence of $\xi_{\rm loc}$ with the dimerization, for $\delta_{N/2}/t_{\rm C} = 0.1, 0.5,8$. The dashed line corresponds to Eq. (\ref{generalized.dimerization}).}
		\label{figure.dimerization.and.winding.number}}
\end{figure}

\emph{Study in the one-excitation subspace.-} 
The preparation of single excitations can be easily realized in trapped-ion chains, as demonstrated in implementations of the Ising and XY models \cite{Jurcevic2014N,Richerme2014N}.
The one-particle sector is spanned by the vectors $\left| j \right\rangle \equiv \sigma^+_j \left| \downarrow\downarrow \downarrow\cdots\right\rangle,j=1,\ldots,N$. We can think of the state $\left| \downarrow\downarrow \downarrow\cdots\right\rangle$ as a vacuum of particles, and accordingly $\left| j \right\rangle$ represents an excitation localized at site $j$. Since (\ref{hamiltonian.ssh.spins.long.range}) is invariant under arbitrary rotations in the $xy$ plane, the Hamiltonian does not mix $\left| j \right\rangle$ with states within subspaces of different number of excitations. Therefore, the dynamics of $\left|j\right\rangle$ is dictated by the restriction of the Hamiltonian to the one-excitation subspace, that is given as
\begin{equation}
\bar{H}^{\rm (ions)}_{\rm SSH}=\sum_{j,l=1}^N {h_{j,l}}\left(|j\rangle \langle l|
+| l\rangle \langle j|\right),\ h_{j,l}=J_{j,l}^{(\rm ions)} \mathcal{J}_{j,l}^{\pi/2}.
\label{hamiltonian.one.body}
\end{equation}
For $\phi=3\pi/4$ and $\eta>0$, $h_{j,l}$ possesses two (quasi-) zero-energy modes, which feature localization at the edges; we show one of these in Fig.\ref{figure.dimerization.and.winding.number}(a).
The edge state has appreciable support only on the odd sites, which is a consequence of the chiral symmetry \cite{Asboth2016}. Indeed, the chiral-symmetric limits of this Hamiltonian are attained for $\phi=\pi/4$ and $3\pi/4$ (see Supplemental Material). We have depicted the dimerization (\ref{generalized.dimerization}) as a function of $\eta$ in these limits (cf. Fig.\ref{figure.dimerization.and.winding.number}(b)). We note that for $\phi = 3\pi/4$, $\delta$ is positive, and accordingly the model presents edge states. This is accompanied by a non-zero value of the associated bulk invariant, the Zak phase \cite{Zak1989PRL}, which can take the value $0(\pm \pi)$ in the trivial (topological) phase. As shown in Fig.\ref{figure.dimerization.and.winding.number}(c), the Zak phase is $0$ or $\pm\pi$ in the chiral limits $\phi = \pi/4$ and $\phi = 3\pi/4$, signaling the emergence of edge states in this latter case.

By fitting the edge state to an exponential, we can estimate its localization length numerically. According to Eq. (\ref{loc.length.discrete}), this quantity is a decreasing function of the dimerization. This holds true for $\bar{H}^{\rm (ions)}_{\rm SSH}$, as shown in Fig.\ref{figure.dimerization.and.winding.number}(d). However, we note that $\xi_{\rm loc}$ decreases with the range of the interactions as well, i.e., \emph{there is an enhancement of the localization effect}. This feature is not captured by the prediction for the original SSH model, since $\xi_{\rm loc}$ exclusively depends on $\delta$, and this latter quantity is insensitive to the range of the couplings (cf. Eq. (\ref{generalized.dimerization})). To obtain the dependence of the localization on the interaction range we have considered the effective theory for the low-energy sector of $\bar{H}^{\rm (ions)}_{\rm SSH}$, which captures the long-range effects by a renormalization of the parameters of the theory compared to those of the original SSH model.

\begin{figure}[h!]{
		\centering
		\includegraphics[width=1.\linewidth]{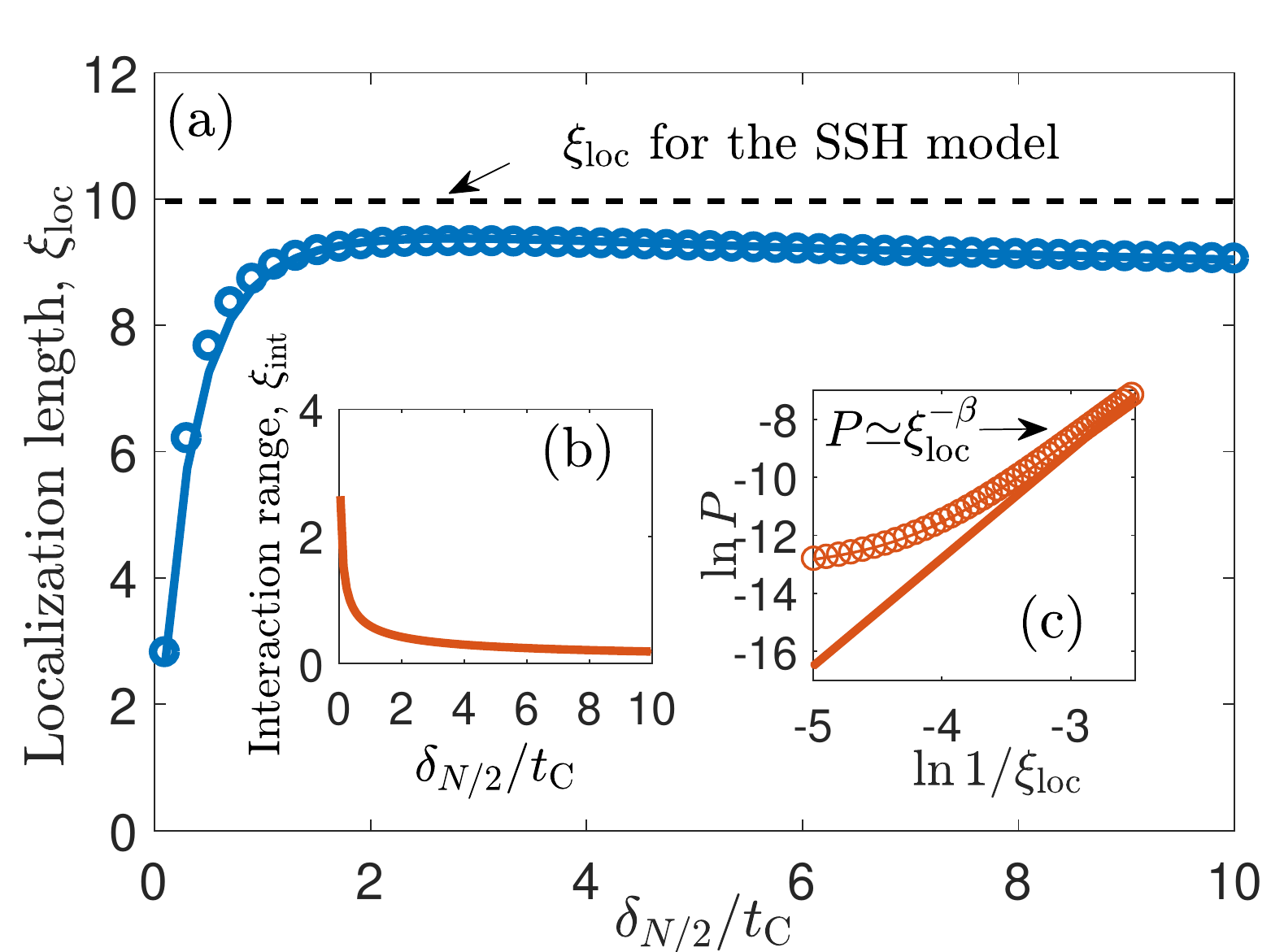}
		\caption{(color online) (a) Localization length of the edge state, from the exact diagonalization of $\bar{H}^{\rm (ions)}_{\rm SSH}$ (solid line) for $N=100$ sites, and from expression (\ref{prediction.loc.length}) (circles) with $\delta = 0.1$ ($\phi = 3\pi/4$, $\eta\simeq 0.62$). The largest enhancement of $\xi_{\rm loc}$ occurs for $\delta_{N/2}/t_{\rm C}<1$. (b) Interaction range, $\xi_{\rm int}$, of the exponentially decaying component of $J_{j,l}^{\rm (ions)}$. (c) Log-log plot of the survival probability $P$ as a function $1/\xi_{\rm loc}$. For $\xi_{\rm loc}\to 1$, $P\sim \xi_{\rm loc}^{-\beta}$ with $\beta\simeq 3.8$, consistent with the prediction (\ref{survival.probability}). We take $\delta_{N/2}/t_{\rm C}=1/3$, $N=1000$, and values of $\eta$ in the interval $0.13-0.5$, for $\phi = 3\pi/4$.}
		\label{figure.localization.vs.detuning}}
\end{figure}

\emph{Localization length of the edge states of $\bar{H}_{\rm SSH}^{\rm (ions)}$.-} 
The effective theory of the SSH model in $k$-space can be described in terms of pairs of states $|k, \pm  \rangle = |k \pm k_{\rm F}\rangle$, where $k_{\rm F}\equiv\pi/2$. The low energy Hamiltonian is given by $H_{\rm low-E} = (N/2\pi)\int_{-\pi/2}^{\pi/2} h(k) dk$, with
\begin{eqnarray}
h(k) 
&=& k v_{\rm F} \left( 
|k,+\rangle \langle k,+| \ -  \
|k,-\rangle \langle k,-| 
\right)
\nonumber \\ 
&-& 
 i \ \Delta_0 \left( |k,+\rangle \langle k,-| \ - \ |k,-\rangle \langle k,+| \rangle\right).
\end{eqnarray}
The two parameters $v_{\rm F}=2J$ and $\Delta_0=2J\delta$ fully characterize 
the low energy-sector. From them, the dimerization is directly obtained 
as $\Delta_0/v_{\rm F}=\delta$, and since the effective theory assumes that 
$\Delta_0\ll v_{\rm F}$, we can approximate the localization length 
(\ref{loc.length.discrete}) as
\begin{equation}
\xi_{\rm loc} \sim \frac{v_{\rm F}}{\Delta_0}.
\label{prediction.loc.length}
\end{equation} 
This prediction must hold for any lattice model whose low-energy excitations are captured by a Hamiltonian such as $H_{\rm low-E}$. In particular, this is the case for $\bar{H}^{\rm (ions)}_{\rm SSH}$, that can be rewritten as $\sum_{j=1}^{N} \sum_{d=1-j}^{N-j} h^{(d)}_j \left(|j\rangle \langle j+d|+| j+d\rangle \langle j|\right)$, where $h^{(d)}_j \equiv J_d^{\rm (ions)} (\mathcal{J}^{(+)}_d + \mathcal{J}^{(-)}_d (-1)^j)$, with
\begin{equation}
\mathcal{J}^{(\pm)}_d = \frac{1}{2}\left(\mathcal{J}^{\rm even}_d\pm \mathcal{J}^{\rm odd}_d\right),
\end{equation} 
and the latter quantities defined as $\mathcal{J}^{\pi/2}_{j,j+d}$ for $j$ even or odd, respectively. In terms of plane waves, and assuming $N\to \infty$, we obtain
\begin{equation}
\bar{H}_{\rm SSH}^{(\rm ions)}=\sum_k \epsilon'(k) \left|k\right\rangle \left\langle k\right| +\sum_{k} \Delta'(k) \left|k+\pi \right\rangle \left\langle k\right|+{\rm H.c.},
\end{equation}
where we have defined 
\begin{equation}
\left\{\begin{array}{lcl}
\epsilon'(k)&=&\displaystyle 4\sum_{d=1}^{\infty} J_d^{\rm (ions)}\mathcal{J}^{(+)}_{d} \cos{\left(k d\right)},\\
\Delta'(k)&=& \displaystyle 2\sum_{d=1}^{\infty} J_d^{\rm (ions)}\mathcal{J}^{(-)}_{d} e^{ikd}.
\end{array}\right.
\end{equation}
From these quantities, we can obtain the parameters of the effective theory as (see Supplemental Material)
\begin{equation}
v_{\rm F}' = \left.\frac{\partial\epsilon'(k)}{\partial k}\right|_{k=k_{\rm F}},\quad \Delta_0' = 2\operatorname{Im}\left(\Delta'(k=k_{\rm F})\right),
\end{equation} 
and compute the localization length (\ref{prediction.loc.length}). We show that this prediction accurately holds for several values of $\delta_{N/2}/t_{\rm C}$ in Fig.\ref{figure.localization.vs.detuning}a, along with the corresponding interaction range (cf. Fig.\ref{figure.localization.vs.detuning}b).

The localization enhancement could be actually measured in an experiment. The idea is to unveil the existence of the edge state by studying the dynamics of a single excitation at the boundary \cite{Jurcevic2014N,Richerme2014N}. To detect an edge state located at, e.g., the left end of the chain, we can prepare the `excited state' $|\psi(t=0)\rangle=\left|\uparrow\downarrow\downarrow\ldots\right\rangle$, which has a large overlap with the boundary mode, and look at its survival probability at long times, $P\equiv |\langle \psi(t)|\sigma^+_1\sigma_1^-|\psi(t)\rangle|^2,t\to\infty$. This quantity can be estimated as (see Supplemental Material)
\begin{equation}
P\left(\frac{1}{\xi_{\rm loc}}\right) \simeq \left(\frac{c_1}{\xi_{\rm loc}^2}+\frac{c_2}{N}\right)^2.
\label{survival.probability}
\end{equation}
Since the overlap is appreciable only if the Hamiltonian presents an edge state, $P$ will take negligible values except in the event of localization at the left end. The initial condition $|\psi(t=0)\rangle$ requires applying a $\pi$ pulse to the leftmost ion in the chain, which in turn can be prepared in the `ground state' $\left|\downarrow\downarrow\downarrow\ldots\right\rangle$ by optical pumping \cite{Schneider2012RoPiP}. Then we can switch on the Hamiltonian $H_{\rm SSH}^{(\rm ions)}$, and wait up to $t\gg \Delta_0^{-1}$, where $\Delta_0$ is the lowest energy scale in the Hamiltonian. Finally, we can perform a fluorescence measurement of the state of the leftmost ion. 
We have numerically confirmed the dependence of $P$ on $\xi_{\rm loc}$ (cf. Eq. (\ref{survival.probability})) in Fig.\ref{figure.localization.vs.detuning}c. Deviations from the power law $P\simeq \xi_{\rm loc}^{-\beta}$, with $\beta = 4$, are the consequence of finite size effects, which play a less important role when $1/\xi_{\rm loc}\ll N$. 

\begin{figure}[h!]
		{\centering
		\includegraphics[width=0.9\linewidth]{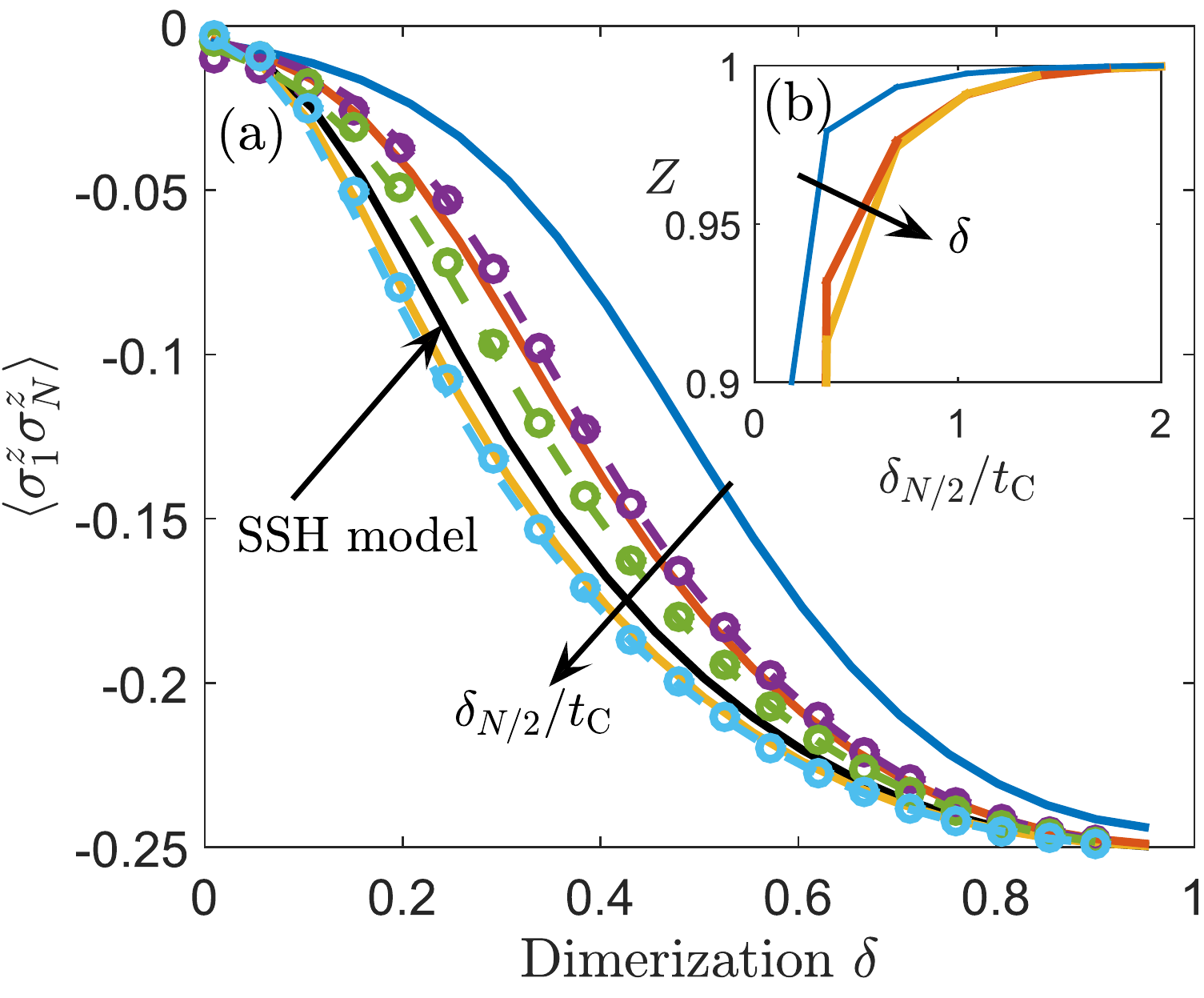}
        \caption{(color online) (a) Correlations $\langle \sigma^z_1 \sigma^z_N \rangle$ for $N=16$, $\phi = 3\pi/4$ and using $\eta$ to tune the dimerization. The arrow shows the direction of decreasing range of the interactions, or increasing detuning from the bottom of the motional band. We have plotted curves for $\delta_{N/2}/t_{\rm C}=0.5,1$ and $10$. The solid lines represent the exact result from Hamiltonian (\ref{hamiltonian.ssh.spins.long.range}), the dashed lines the results from the truncated Hamiltonian (see Supplemental Material), and the circles correspond to the predictions of the HF approximation. (b) Value of the parameter $Z$ for different interaction ranges and $\delta = 0.1,0.3,0.5$ from bottom to top.}
        \label{figure.correlations.vs.dimerization}
        }
\end{figure}

\emph{Correlations in the many-body ground-state.-} 
So far we have been dealing with the single-excitation subspace. Nevertheless, we expect 
that some localization at the edges features as well in the ground state of the many-body 
Hamiltonian (\ref{hamiltonian.ssh.spins.long.range}). 
In a finite chain, states localized at each end hybridize to give rise to solutions that have support at the left and right boundaries. 
We expect that the correlations between the ends are zero if there is no localization at the edges whereas they must have a non-zero value otherwise, a result that has been established for the SSH model \cite{CamposVenuti2007PRA}. 
We illustrate this fact in Fig. \ref{figure.correlations.vs.dimerization}, where we have 
computed $\langle \sigma^z_1 \sigma^z_N\rangle$ as a function of the dimerization for both 
$H_{\rm SSH}$ and $H_{\rm SSH}^{(\rm ions)}$. 
The correlations in the original SSH model are non-zero for $\delta>0$ as expected. This holds qualitatively true for $H_{\rm SSH}^{(\rm ions)}$ as well. 
Indeed, in the regime of short range of the interactions the correlations are larger than those of the original SSH model, which is consistent with the enhanced localization length predicted in the one-excitation subspace (cf. Fig. \ref{figure.localization.vs.detuning}). Conversely, we observe a degradation of the correlations in the long-range interaction regime, i.e., for $\delta_{N/2}\to 0$ there is a decrease in the localization effect. 
This result is a consequence of the mixing --induced by the interactions-- of the 
single-particle edge states with the bulk modes. To quantify this effect we express our generalized SSH model in terms of Jordan-Wigner fermions as
%
$H_{\rm SSH}^{(\rm ions)} = \sum_{l> j}^N 2 J_{j,l}^{(\rm ions)} \mathcal{J}_{j,l}^{\pi/2}\left(c^{\dagger}_jK_{j,l}c_l
+c_jK_{j,l} c_l^{\dagger}\right)$,
%
where $K_{j,l} \equiv \prod_{m=j}^{l-1}(1-2c^{\dagger}_m c_m)$. 
We neglect terms for which $|j-l|\geq 3$ and recast this problem as 
$H_{\rm SSH}^{(\rm ions)} \simeq H_0 + H_{\rm int}$, 
where $H_0 = \sum_{j=1}^N (J_j^{(1)}c^{\dagger}_j c_{j+1} + J_{j}^{(2)}c^{\dagger}_j c_{j+2}+{\rm H.c.})$, with $J_j^{(\alpha)} = 2 J^{\rm (ions)}_{j,j+\alpha} \mathcal{J}^{\pi/2}_{j,j+\alpha}$ and
\begin{equation}
H_{\rm int} = -2\sum_{j=1}^N J^{(2)}_j(c^{\dagger}_j c^{\dagger}_{j+1}c_{j+1} c_{j+2} +{\rm H.c.}).
\end{equation}
We deal with the interaction term within the Hartree-Fock approximation \cite{Altland2010}, which renders a simplified Hamiltonian quadratic in fermion operators (see Supplemental Material)
\begin{equation}
H_{\rm HF} = \sum_{\mu=1}^N \epsilon_{\mu} c^{\dagger}_{\mu} c_{\mu} -2 \sum_{{\mu},{\mu}'=1}^N V_{{\mu},{\mu}'} c^{\dagger}_{\mu} c_{{\mu}'}.
\end{equation}
$H_{\rm HF}$ is expressed in terms of the eigenstates of $H_0$, which correspond to the solutions of the Hamiltonian in the one-excitation subspace (cf. Eq. (\ref{hamiltonian.one.body})), that is, $c_j = \sum_{\mu=1}^N M_{j,{\mu}}c_{\mu}$. 
The one-body edge states are eigenstates of $H_0$ and $V_{\mu,\mu'}$ induces the mixing of 
these states with the bulk modes. 
We quantify this effect with a parameter, $Z$, which measures the overlap between the unperturbed boundary modes and the corresponding states in the presence of interaction, and which can be estimated by elementary perturbation theory as $Z\simeq 1 -  \sum_{{\mu}\neq\rm E.S.}^N {4|V_{E.S.,{\mu}}|^2}/{(\epsilon_{E.S.}-\epsilon_{\mu})^2}$. We show this quantity as a function of the range of interactions in the inset of Fig. \ref{figure.correlations.vs.dimerization}. Accordingly, when $\delta_{N/2}\to 0$ the fidelity drops significantly, signaling the decay of the edge modes into the continuum of the states in the bulk. 
Finally the average $\langle \sigma^z_1 \sigma^z_N \rangle$ can be measured in an experiment by detecting the photo-luminescence from individual ions at the ends of the chain (e.g. by electron-shelving techniques \cite{Leibfried2003RMP}).

\emph{Conclusions and outlook.-} In this work we have established the feasibility of implementing a topological insulator with trapped-ion quantum matter. We have shown that the edge states get more localized because of the long-range interactions in ion chains, and that the localized solutions survive to the interactions in the many-body ground state. An immediate extension of this work would consist in the computation of the Zak phase of the many-body ground state, and establishing the symmetries of the model, to shed light on a prospective bulk-edge correspondence in this system. Our ideas could be extended to systems of cold atoms \cite{Bloch2012NP,Ruostekoski08pra} or superconducting qubits \cite{Dalmonte2015PRB}, where dipolar interactions are available.

\acknowledgements \emph{Acknowledgements.-} We are greatly indebted to Alejandro Berm\'udez for very useful discussions. P.N. thanks Hiroki Takahashi for his feedback on the implementation. Our research has received funding from the People Programme (Marie Curie Actions) of the European Union’s Seventh Framework Programme (FP7/2007-2013) under REA Grant Agreement No. PCIG14-GA-2013-630955.

\appendix

\section{Appendix A: Trapped-ion experimental parameters} 
We consider a chain of $N$ ions along the trap axis, $z$, and we focus on their 
displacements in the transversal direction, $x$. 
The bare dynamics of the ions is described 
by the Hamiltonian 
\begin{equation}
H_{0} = H_{\rm ph} + H_{\rm s}=\sum_{n=1}^N\omega_{n} a_{n}^{\dagger}a_{n} 
+ \frac{\omega_0}{2} \sum_{j=1}^N \sigma^z_j .
\end{equation}
$\omega_0$ is the frequency of the electronic transition, and 
$\omega_{n}$ are the frequencies of the normal modes of motion of the chain, with creation 
(annihilation) operators $a^{\dagger}_{n}(a_{n})$. A pair of laser beams transverse to the 
chain can drive optical Raman transitions leading to a spin-dependent force, whose 
Hamiltonian in the interaction picture is given as 
\cite{Sorensen1999PRL,Solano1999PRA,Sackett2000N,Lee2005JOB}
\begin{equation}
{H}_{\rm f}(t) = g \sum_{j,n=1}^N \sigma^{x}_j(M_{j,n}e^{i\delta_nt} a_{n}^{\dagger} +{\rm H.c.}),
\label{spin.phonon.hamiltonian.zforce}
\end{equation}
with $\delta_n \equiv \omega_{n} - \Delta \omega$, where $\Delta\omega$ is the laser 
detuning from the internal transition frequency $\omega_0$. 
It is customary to work in a 
modified interaction picture, in which phonons rotate with frequency $\Delta \omega$, so that ${H}_{\rm f}(t) \to {H}_{\rm f}(0)$ \cite{Schneider2012RoPiP}. In this picture, the energies of the phonon modes get shifted to $\omega_{n} - \Delta\omega=\delta_n$, so that as long as $g\ll\delta_n$ the phonons are only virtually excited. 
In this regime, the phonon and the spin degrees of freedom decouple, and the latter obey the phonon-mediated Ising-type interaction term in Hamiltonian (\ref{hamiltonian.ising}).
The specific form of those Ising interactions can be analytically estimated in the limit $N \gg 1$,
\begin{eqnarray}
J_{j,l}^{\rm (ions)} &=& - \sum_n M_{j,n} M_{l,n} \frac{g^2}{\delta_n} 
\nonumber \\
&=& -(-1)^{(j-l)} J_{\rm exp} e^{-|j-l|/\xi_{\rm int}} +
\frac{J_{\rm dip}}{|j-l|^3},
\label{spin.interaction}
\end{eqnarray} 
with the constants
\begin{eqnarray}
\xi_{\rm int} &=& \sqrt{\frac{\ln 2}{2}} \sqrt{\frac{t_{\rm C}}{\delta_{N/2}}} , \nonumber \\
J_{\rm exp} &=& \frac{\xi_{\rm int} g^2}{t_{\rm C} \ln 2} , \nonumber \\
J_{\rm dip} &=& \frac{g^2 t_{\rm C}}{2 \left(\delta_{N/2} + 7 \zeta(3) t_{\rm C} /4 \right)^2} .
\label{parameters.interaction}
\end{eqnarray}
A detailed derivation of the effective spin-spin interaction in Eqs. (\ref{spin.interaction}, \ref{parameters.interaction}) was carried out in our Ref.  \cite{Nevado2016PRA}, but the form of the coupling has a very clear physical interpretation.
In particular, in the case $\delta_{N/2} \ll t_{\rm C}$ the dominant contribution is an exponential decay of the spin couplings which has the form of the usual Yukawa coupling mediated by a bosonic field. The range of the couplings, $\xi_{\rm int}$, scales like $1/\sqrt{\delta_{N/2}}$, since $\delta_{N/2}$ is the minimum vibrational energy (energy gap). In the opposite limit, $\delta_{N/2} \ll t_{\rm C}$ we recover a dipolar decay of the spin-spin interaction which to all effects is equivalent to a short-range interaction
\cite{Deng2005PRA}.

The Hamiltonian for the simulation of the SSH is the sum of two terms, $H_{\rm Ising}$, Eq. (\ref{hamiltonian.ising}) and the driving term $H_{\rm driving}$ in Eq. (\ref{H.driving}). We discuss below what are the physical values of the parameters governing those Hamiltonian terms in a realistic trapped ion experiment. 

For the implementation of the spin-spin interaction in Eq. (\ref{hamiltonian.ising}) we consider a couple of lasers inducing a spin-dependent force upon the transverse modes. 
For concreteness, we assume a (homogeneous) crystal of $^9\rm{Be}^+$ ions along a Paul trap, such that each ion hosts a hyperfine qubit. However, our ideas can be also implemented with optical transitions, for example in $^{40}\rm{Ca}^+$ ions, where spin-dependent forces and effective spin-spin interactions have been demonstrated \cite{Blatt2012NP}. Ions in the chain are separated by distances $d_0=10\,\mu$m, with a transversal trapping frequency,  $\omega_x = 5 (2\pi)$ MHz.
In this limit, the spectral width of the radial modes is $t_{\rm C} \simeq 77.2\,(2\pi)$ kHz. In the short-range limit of the effective couplings, we have that $J_{j,l}^{\rm (ions)} \simeq {J/}{|j-l|^3}, \text{ with } J \simeq {g^2 t_{\rm C}}/{2(\delta_{N/2})^2}$, where $\delta_{N/2}$ is the detuning from the bottom of the radial modes dispersion relation. We assume that $\delta_{N/2}\simeq 2g$, so that we estimate $J\simeq 10\,(2\pi)$ kHz. This quantity determines the slowest frequency in the experiment. 
Typical magnitudes of the effective Rabi frequency are $\Omega = 100\, (2\pi)$ kHz, or even higher \cite{Leibfried2003RMP}. We see that the condition $\Omega \gg J^{\rm ions}_{j,l}$ that is required to eliminate fast rotating terms of the form 
$\sigma^+_j \sigma^+_l$, $\sigma^-_j \sigma^+_l$ can be easily satisfied.

The periodic driving needs to be implemented by a different set of lasers than the optical force. Two counter-propagating lasers with wave-vectors ${\bf k}_1$ and ${\bf k}_2$ can induce a standing-wave leading to the coupling in Eq. (\ref{H.driving}). 
The optical phase of the standing-wave, $\phi$, relative to the position of the ions must be stable along the duration of the experiment and it must be adjusted to reach the values considered in the main text. However, we stress that small deviations in the position of the ions relative to the standing-wave should not strongly affect general features observed in the experiments, such as the observation of edge states, which can be observed over a wide range of values of the dimerization parameter, $\delta$.
The wave-vector of the standing wave, 
$\Delta {\bf k} = {\bf k}_2 - {\bf k}_1$ needs to have a component parallel to the trapped ion chain axis that we can express like 
$\Delta k_z = \sin(\theta) |\Delta {\bf k}| = \sin(\theta) \left(2 \pi /\lambda \right)$. A typical value in trapped ion experiments is $\lambda =$ 320 nm. The condition $|\Delta k_z| d_0 = \pi/2$, considered in the main text can be achieved with a standing-wave almost perpendicular to the ion chain and a small tilting angle, $\theta \simeq 0.46$ degrees. Since the periodic driving in Eq. (\ref{H.driving}) is intended to dress the spin-spin interactions and not to induce any further spin-phonon coupling, it is important that it is out of resonance with the vibrational modes of the chain. To ensure this condition, we need to specify the axial trapping frequency, which can be estimated as $\omega_z = 192 \, (2\pi)$ kHz for $N=20$ ions and the average distance $d_0 = 10 \, \mu m$. 
Under those conditions we can choose the value $\omega_{\rm d} \simeq 50$ kHz, so that $\omega_x, \omega_z > \omega_{\rm d}$, and the periodic driving is not able to induce vibrational transitions. With the values considered here, the condition $\max_{j,l}|J_{j,l}^{(\rm ions)}|\ll \omega_{\rm d}\ll \Omega$ required to derive our generalized SSH Hamiltonian is also satisfied. 

Finally, we recall that the typical time associated with the detection of the edge state is $\Delta_0^{-1}$. Since $\Delta_0\simeq 2J\delta$, for $\delta=0.1$ ($\eta\simeq0.62,\phi=\pi/4$), we have that $\Delta_0^{-1}\simeq 0.17$ ms. This is consistent with experimental times for the preparation and detection of many-body spin states in trapped-ion quantum simulators \cite{Kim2010N}.

\section{Appendix B: Chiral limits of $\bar{H}^{\rm (ions)}_{\rm SSH}$}

To discuss the chiral symmetry of (\ref{hamiltonian.one.body}), we write this problem in a fashion that highlights the two-fold periodicity of the couplings (\ref{bessel.couplings}). The terms $\mathcal{J}_{j,l}^{\pi/2}$ naturally belong to one of two possible sublattices, which are comprised by the odd ($A$) and even ($B$) sites. Thus, the chain is made up by $n=1,\ldots,N/2\equiv M$ dimers, each of them consisting of two adjacent sites of sublattices $A$ and $B$, and we can express (\ref{hamiltonian.one.body}) as
\begin{eqnarray}
\bar{H}_{\rm SSH}^{\rm (ions)} &=&
\sum_{n,m=1}^{M} J^{AA}_{n,m}\mathcal{J}^{AA}_{n,m} \left|n,A\right\rangle \left\langle m,A\right|\nonumber\\
&+& \sum_{n,m=1}^{M} J^{BB}_{n,m}\mathcal{J}^{BB}_{n,m} \left|n,B\right\rangle \left\langle m,B\right| \nonumber\\
&+&J^{AB}_{n,m}\sum_{n,m=1}^{M}\left(\mathcal{J}^{AB}_{n,m} \left|n,A\right\rangle \left\langle m,B\right|\right.\nonumber\\
&+&\left.\mathcal{J}^{BA}_{n,m} \left|n,B\right\rangle \left\langle m,A\right|\right),
\end{eqnarray}
where the coefficients are the corresponding restrictions of $J^{\rm(ions)}_{j,l}$ and $\mathcal{J}^{\pi/2}_{j,l}$ to sublattices $A$ and $B$. Furthermore, $J^{AA}_{n,m}=J^{BB}_{n,m}$. Now we make a transformation of the `external' degrees of freedom $\left|n\right\rangle$ into the plane wave basis $\left|\mu\right\rangle = \sum_{n=1}^M e^{i\frac{2\pi n}{M}\mu}/\sqrt{M} \left|n\right\rangle$, so $\bar{H}_{\rm SSH}^{\rm (ions)} = \sum_{\mu=0}^{M-1} h_{\mu}\left|\mu\right\rangle \left\langle \mu \right|$, with
\begin{equation}
h_{\mu}=\begin{pmatrix}
\sum_d J^{AA}_d\mathcal{J}^{AA}_{d} e^{i\frac{2\pi d}{M}\mu} & \sum_d J_d^{AB}\mathcal{J}^{AB}_{d} e^{i\frac{2\pi d}{M}\mu}\\[1em]
\sum_d J_d^{BA}\mathcal{J}^{BA}_{d} e^{-i\frac{2\pi d}{M}\mu} & \sum_d J_d^{BB}\mathcal{J}^{BB}_{d} e^{i\frac{2\pi d}{M}\mu} \\
\end{pmatrix},
\label{pseudospin.rep}
\end{equation}
where the sums run from $d=0$ to $d=M-1$, and $d\equiv n-m$. We can associate a vector $\mathbf{d}_{\mu}\in \mathbb{R}^3$ with $h_{\mu}$ through the identification $h_{\mu}=d_{\mu}^0\sigma^0+ \mathbf{d}_{\mu}\cdot\bm{\sigma}$, where $\sigma^0$ is the $2\times2$ identity matrix, and $\bm{\sigma}=\left(\sigma^x,\sigma^y,\sigma^z\right)$. The chiral symmetry is attained in the event of $\sigma^z \left(\mathbf{d}_{\mu}\cdot\bm{\sigma}\right) \sigma^z = -\mathbf{d}_{\mu}\cdot\bm{\sigma}$ \cite{Asboth2016}, which entails that $\mathcal{J}^{AA}_{d}=\mathcal{J}^{BB}_{d}$. This boils down to the condition
\begin{equation}
\sin \left(\frac{\pi}{2}+\phi\right)=\pm\cos \left(\frac{\pi}{2}+\phi\right),\phi\in [0,\pi],
\end{equation}
that holds for $\phi = \pi/4,3\pi/4$, and all $d$ and $\eta$. The former values constitute the chiral-symmetric limits of $\bar{H}^{\rm (ions)}_{\rm SSH}$. Regarding the Zak phase, which is given as \cite{Zak1989PRL}
\begin{equation}
\nu = i \oint \langle u(k)| \partial_k |u(k)\rangle dk,
\label{zak.phase}
\end{equation}
we have computed $\nu$ by a discretization of the Brillouin zone, and using the gauge-independent formula $\nu \simeq -\operatorname{Im}\log\prod_{\mu=0}^{M-1}\langle u_{\mu +1}| u_{\mu}\rangle$ \cite{Resta1994RMP}, where $| u_{\mu}\rangle$ is the ground state of $h_{\mu}$, and $|u_0\rangle=|u_M\rangle$.

\section{Appendix C: Continuum theory of the generalized SSH model}

\begin{figure}[h!]
	\centering
	\includegraphics[width=1\linewidth]{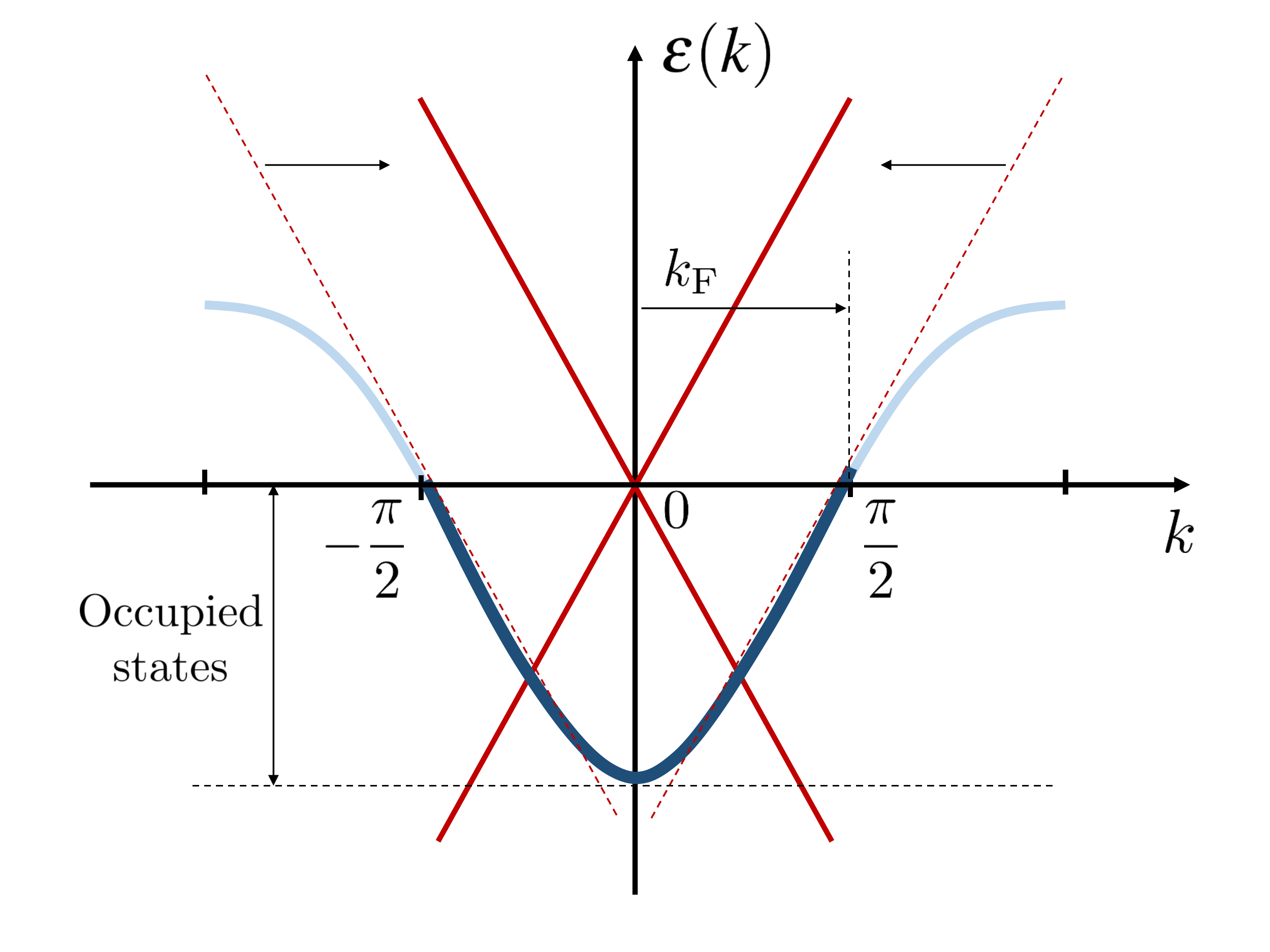}
	\caption[Energy bands of the continuum theory]{(color online)Band diagram of ${H}_{\rm SSH}$ for $\Delta_{\mu} =0$ (sinusoid). The (many-body) ground state is comprised by all the eigenstates of ${H}_{\rm SSH}$ with $\epsilon(k)<0$ (dark blue). The energy bands of the continuum theory (red lines) are linear in $k$, and their slope is set by the derivative of $\epsilon(k)$ at the Fermi points $k=\pm\pi/2$. The low-energy physics occurs just above $\epsilon(k)=0$.}
	\label{diagram.continuum.bands}
\end{figure}

The basic idea for obtaining the effective theory is to consider the dimerization as a perturbation \cite{Takayama1980PRB,Heeger1988RMP}. We illustrate this for the original SSH model following reference \cite{Asboth2016}; the discussion will apply straightforwardly to the generalized model. We assume that we work in the one-excitation subspace, in which Hamiltonian (\ref{h.ssh.original}) reduces to $\bar{H}_{\rm SSH} = \sum_{j=1}^{N} h_j \left(|j\rangle \langle j+1|+| j+1\rangle \langle j|\right)$, where $h_j = J(1+(-1)^j)$. By transforming $\bar{H}_{\rm SSH}$ into the plane-wave basis, and taking the limit $N\to \infty$, we arrive at
\begin{equation}
\bar{H}_{\rm SSH}=\sum_{k} \epsilon(k) \left|k\rangle \langle k\right| +\sum_{k} {\Delta(k)} \left|k\rangle \langle k+\pi\right|+{\rm H.c.},
\label{app.one.ex.hamiltonian}
\end{equation}
where $k\in [0,2\pi]$, and we have introduced the band dispersion relation $\epsilon(k) = 2 J \cos\left(k\right)$, and the scattering potential $\Delta(k) = -J\delta e^{-ik}$. We note that, for a given $k$, the Hamiltonian mixes momenta $\left|k\right\rangle$ and $\left|k+\pi\right\rangle$ exclusively. Therefore, we can constraint the sum in (\ref{app.one.ex.hamiltonian}) to the interval $k\in[0,\pi]$, and think of two different kind of excitations, $\left|k\right\rangle\equiv\left|k,+\right\rangle$ and $\left|k+\pi\right\rangle\equiv\left|k,-\right\rangle$. In terms of these, $\bar{H}_{\rm SSH}$ can be rewritten as 
\begin{eqnarray}
\bar{H}_{\rm SSH} 
&=& \sum_k \epsilon(k) \left( 
|k,+\rangle \langle k,+| \ -  \
|k,-\rangle \langle k,-| 
\right)
\nonumber \\ 
&-& 
2i \ \operatorname{Im}(\Delta(k)) \left( |k,+\rangle \langle k,-| \ - \ |k,-\rangle \langle k,+| \rangle\right).
\end{eqnarray}
%
To isolate the low-energy excitations upon the correct many-body ground state, we must assume that $\bar{H}_{\rm SSH}$ is comprised of fermionic excitations. Also, for later convenience, we shift the momenta $k$ to the interval $[-\pi/2,\pi/2]$. Then, as long as $|\Delta(k)|\ll \epsilon(k),\forall k$, the low-energy processes occur in an energy window of width $\sim |\Delta(k)|$ around the Fermi level. Thus, it is justified to approximate the dispersion relation by its derivative at the Fermi momenta $k=\pm k_{\rm F}$, which correspond to $k_{\rm F} \equiv \pi/2$ (cf. Fig. \ref{diagram.continuum.bands}).
This boils down to assume that $\epsilon(k\pm k_{\rm F}) = 2J\cos(k\pm\pi/2) \simeq \pm 2Jk\equiv \pm v_{\rm F} k$. Because of the degeneracy of the states at $k=\pm k_{\rm F}$, the first-order processes induced by the dimerization are necessarily scattering events between them --therefore lifting their degeneracy--, and their magnitude is given as $\pm \Delta_0 \equiv 2\operatorname{Im}\left(\Delta(k=\pm k_{\rm F})\right)= \pm 2J\delta$. Thus, the theory of the low-energy sector can be written as $H_{\rm low-E} = (N/2\pi)\int_{-\pi/2}^{\pi/2} h(k) dk$, with
\begin{eqnarray}
h(k) 
&=& k v_{\rm F} \left( 
|k,+\rangle \langle k,+| \ -  \
|k,-\rangle \langle k,-| 
\right)
\nonumber \\ 
&-& 
i \ \Delta_0 \left( |k,+\rangle \langle k,-| \ - \ |k,-\rangle \langle k,+| \rangle\right).
\end{eqnarray}
This form is the same for both the SSH and the generalized models, the only difference being the particular values of the parameters $v_{\rm F}$ and $\Delta_0$.

\section{Appendix D: Survival probability of an excitation at the edge}

The dynamics of the state $\left|\uparrow\downarrow\downarrow\ldots\right\rangle$ is dictated by Hamiltonian (\ref{hamiltonian.one.body}), so the corresponding
Schr{\"o}dinger equation for $|\psi(t)\rangle = \sum_{j=1}^Nc_j(t)|j\rangle$ reads $i\dot{c}_j(t)= 2\sum_{l=1}^N h_{j,l} c_{l}(t)$, and its solution can be straightforwardly computed as $c_j(t)= \sum_{n,j'=1}^{N} e^{-i 2\epsilon_n t}M_{j,n} M_{j',n} c_{j'}(0)$, where $c_j(0)=\delta_{1,j}$, and $\epsilon_n$ and $M_{j,n}$ are the eigenvalues and eigenstates of $h_{j,l}$. The probability amplitude that the initial state does not diffuse into the bulk for long times is straightforwardly computed as $\langle \psi(t)|\sigma^+_1\sigma_1^-|\psi(t)\rangle\simeq \sum_{n=1}^N |M_{1,n}|^4 $.
%
%
%
%
We expect that the only contribution in the latter sum that depends on the localization length stems from the edge state. This dependence can be estimated by taking into account the normalization of its eigenfunction, which is given as $M_{j,n_0} = Z^{-1} e^{(N-j+1)/\xi_{\rm loc}}$. $Z$ can be computed from the condition $\sum_{j=1}^N |M_{j,n_0}|^2=1$, and we obtain that $Z\simeq \xi_{\rm loc}^{-1/2}$ for $\xi_{\rm loc}\gg 1$. Thus, $|M_{1,n_0}|^4 \propto \xi_{\rm loc}^{-2}$. On the other hand, the rest of the states appearing in $P$ contribute each with $1/\sqrt{N}$, so that $P$ is given by (\ref{survival.probability}).

\section{Appendix E: Hartree-Fock theory}

To study the effect of the interactions upon the many-body ground state, we are going to rely on the Hartree-Fock approximation \cite{Altland2010}. To begin with, we write the generalized SSH Hamiltonian in terms of Jordan-Wigner fermions \cite{Lieb1961AoP} as
\begin{equation}
H_{\rm SSH}^{(\rm ions)} = \sum_{l> j}^N 2 J_{j,l}^{(\rm ions)} \mathcal{J}_{j,l}^{\pi/2}\left(c^{\dagger}_jK_{j,l}c_l
+c_jK_{j,l} c_l^{\dagger}\right),
\label{hamiltonian.ssh.fermionic}
\end{equation}
where we have defined $K_{j,l} \equiv \prod_{m=j}^{l-1}(1-2c^{\dagger}_m c_m)$. Since $J_{j,l}^{(\rm ions)}$ decay with the distance, $K_{j,l}$ with $|j-l|\gg 1$ can be neglected on a first approximation. For example, if we truncate the terms for which $|j-l|\geq 3$, this problem can be recast as $H^{\rm(ions)}_{\rm SSH}\simeq H_{\rm trunc}$, with
\begin{eqnarray}
H_{\rm trunc} &=& \sum_{j=1}^N J^{(1)}_j (c^{\dagger}_j c_{j+1}+{\rm H.c.})\nonumber\\
&&+ \sum_{j=1}^N J_j^{(2)} (c^{\dagger}_j(1-2c ^{\dagger}_{j+1}c_{j+1}) c_{j+2}+{\rm H.c.}).
\end{eqnarray}
Now we write $H_{\rm trunc} = H_0 + H_{\rm int}$, where 
\begin{equation}
H_0=\sum_{j=1}^N (J_j^{(1)}c^{\dagger}_j c_{j+1} + J_{j}^{(2)}c^{\dagger}_j c_{j+2}+{\rm H.c.}),
\end{equation}
with $J_j^{(\alpha)} = 2 J^{\rm (ions)}_{j,j+\alpha} \mathcal{J}^{\pi/2}_{j,j+\alpha}$ and
\begin{equation}
H_{\rm int} = -2\sum_{j=1}^N J^{(2)}_j(c^{\dagger}_j c^{\dagger}_{j+1}c_{j+1} c_{j+2} +{\rm H.c.}).
\end{equation}
%
%
We assume that we can diagonalize $H_0$, that is, we can write
\begin{equation}\label{hamiltonian0.diagonal}
H_0 = \sum_{\mu=1}^N \epsilon_{\mu} c^{\dagger}_{\mu} c_{\mu},\,\text{with } c_j = \sum_{\mu=1}^N M_{j,{\mu}}c_{\mu},\,M_{j,{\mu}}\in \mathbb{R}.
\end{equation}
In terms of the new operators $c_{\mu}$, the interaction term reads
\begin{equation}
\label{hamiltonian.int.diagonal.basis}
H_{\rm int} = -2\sum_{{\mu}_1,{\mu}_2,{\mu}_3,{\mu}_4=1}^N U_{{\mu}_1,{\mu}_2,{\mu}_3,{\mu}_4} c^{\dagger}_{{\mu}_1} c^{\dagger}_{{\mu}_2}c_{{\mu}_3} c_{{\mu}_4},
\end{equation}
where
\begin{eqnarray}
U_{{\mu}_1,{\mu}_2,{\mu}_3,{\mu}_4} &\equiv& \sum_{j=1}^N J^{(2)}_j (M_{j,{\mu}_1}M_{j+1,{\mu}_2}M_{j+1,{\mu}_3}M_{j+2,{\mu}_4}\nonumber\\
&&+M_{j+2,{\mu}_1}M_{j+1,{\mu}_2}M_{j+1,{\mu}_3}M_{j,{\mu}_4}).
\end{eqnarray}
So far, we have not made any approximation. However, the interaction term is difficult to deal with in general, so we rely on the following procedure, which we refer to as the Hartree-Fock approximation: we form all the possible pairings of two operators $c^{\dagger}_{\mu} c_{{\mu}'}$ in $H_{\rm int}$, and evaluate them upon the ground state of $H_{0}$. The other two remaining operators are left unevaluated, and everything is placed in normal order. There are four different pairings possible, e.g.,
$$
\contraction[2ex]{}{c}{{}^{\dagger}_{{\mu}_1} c^{\dagger}_{{\mu}_2}}{c}
c^{\dagger}_{{\mu}_1} c^{\dagger}_{{\mu}_2}c_{{\mu}_3} c_{{\mu}_4},
\contraction[2ex]{}{c}{{}^{\dagger}_{{\mu}_1} c^{\dagger}_{{\mu}_2}c_{{\mu}_3}}{c}
c^{\dagger}_{{\mu}_1} c^{\dagger}_{{\mu}_2}c_{{\mu}_3} c_{{\mu}_4},
\contraction[2ex]{c^{\dagger}_{{\mu}_1}}{c}{{}^{\dagger}_{{\mu}_2}}{c}
c^{\dagger}_{{\mu}_1} c^{\dagger}_{{\mu}_2}c_{{\mu}_3} c_{{\mu}_4} \text{ and } 
\contraction[2ex]{c^{\dagger}_{{\mu}_1}}{c}{{}^{\dagger}_{{\mu}_2}c_{{\mu}_3}}{c}
c^{\dagger}_{{\mu}_1} c^{\dagger}_{{\mu}_2}c_{{\mu}_3} c_{{\mu}_4}.
$$
Since $\langle c^{\dagger}_{\mu} c_{{\mu}'}\rangle = \delta_{{\mu},{\mu}'}$ for ${\mu}\ni \epsilon_{{\mu}}< 0$, we can compute straightforwardly the sums in $H_{\rm int}$, in terms of which we define the Hartree-Fock Hamiltonian
\begin{equation}
H_{\rm HF} = \sum_{\mu=1}^N \epsilon_{\mu} c^{\dagger}_{\mu} c_{\mu} -2 \sum_{{\mu},{\mu}'=1}^N V_{{\mu},{\mu}'} c^{\dagger}_{\mu} c_{{\mu}'},
\end{equation}
where $V_{{\mu},{\mu}'}$ is
\begin{equation}
\sum_{q\ni \epsilon_{q}< 0}^N \left(-U_{q,{\mu},q,{\mu}'}+U_{q,{\mu},{\mu}',q}+U_{{\mu},q,q,{\mu}'}-U_{{\mu},q,{\mu}',q}\right).
\end{equation}
Now we can transform $H_{\rm HF} $ back to the `real space' operators $c_j$, and compute the ground state to check for the correlations.
%
On the other hand, $H_{\rm HF}$ is expressed in terms of the eigenstates of $H_0$, which correspond to the solutions of the Hamiltonian in the one-excitation subspace (cf. Eq. (\ref{hamiltonian.one.body})), that is, $c_j = \sum_{\mu=1}^N M_{j,{\mu}}c_{\mu}$. The one-body edge states are, in particular, eigenstates of $H_0$, and the effective potential $V_{\mu,\mu'}$ induces the mixing of these states with the bulk modes. By use of elementary perturbation theory, the probability for the perturbed edge state to be in any of the eigenstates of $H_0$ can be estimated as $Z\simeq 1 -  \sum_{{\mu}\neq\rm E.S.}^N {4|V_{E.S.,{\mu}}|^2}/{(\epsilon_{E.S.}-\epsilon_{\mu})^2}$ \cite{Sakurai2010}. This is the quantity plotted in Fig.\ref{figure.correlations.vs.dimerization}(b).

\end{document}